\documentclass[12pt]{iopart}
\usepackage{graphicx}
\begin{document}

\title[Post-processing of polymer foam tissue scaffolds with high power ultrasound]{Post-processing of polymer foam tissue scaffolds with high power ultrasound: a route to increased pore interconnectivity, pore size and fluid transport}

\author{N J Watson$^1$, R K Johal$^2$, Y Reinwald$^3$, L J White$^3$, A M Ghaemmaghami$^2$, S P Morgan$^4$, F R A J Rose$^3$, M J W Povey$^1$ and N G Parker$^{1,5}$}

\address{${^1}$School of Food Science and Nutrition, University of Leeds, Leeds, LS2 9JT, UK
\\
$^2$Division of Immunology, School of Molecular Medical Sciences, Queen's Medical Centre, University of Nottingham, Nottingham, NG7 2UH, UK
\\
$^3$School of Pharmacy, Centre for Biomolecular Sciences, University of Nottingham, Nottingham, NG7 2RD, UK
\\
$^4$ Electrical Systems and Optics Research Division, Faculty of Engineering, University of Nottingham,  Nottingham, NG7 2RD, UK
\\
$^5$ School of Mathematics and Statistics, Newcastle University, Newcastle upon Tyne, NE1 7RU, UK}
\ead{nick.parker@ncl.ac.uk}
\begin{abstract}
We expose thick polymer foam tissue scaffolds to high power ultrasound and study its effect on the openness of the pore architecture and fluid transport through the scaffold.  Our analysis is supported by measurements of fluid uptake during insonification and imaging of the scaffold microstructure via x-ray computed tomography, scanning electron microscopy and acoustic microscopy.  The ultrasonic treatment is found to increase the mean pore size by over $10\%$.  More striking is the improvement in fluid uptake: for scaffolds with only 40\% water uptake via standard immersion techniques, we can routinely achieve full saturation of the scaffold over approximately one hour of exposure. These desirable modifications occur with no loss of scaffold integrity and negligible mass loss, and are optimized when the ultrasound treatment is coupled to a pre-wetting stage with ethanol.  Our findings suggest that high power ultrasound is a highly targetted and efficient means to promote pore interconnectivity and fluid transport in thick foam tissue scaffolds.  
\end{abstract}

\maketitle

\section{Introduction}
Biodegradable polymer foams are of major interest as three-dimensional scaffolds for tissue engineering \cite{ma}.  A three-dimensional pore structure provides a high surface area for cell adhesion, while biodegradation leads to the gradual removal of the artificial scaffold as the native extracellular matrix develops.  In such structures, an “open” pore structure is essential to promote homogeneous tissue growth and efficient transport of waste and nutrients.  

The synthetic polymer poly(lactic acid) (PLA) is commonly used for such scaffolds due to its economy, structural versatility, well characterized and tuneable biodegradation, and its long history of use in the clinic \cite{middleton}.  
Conventional solid state foaming techniques based on gas blowing are not directly tractable for scaffold fabrication since they lead to a  closed pore structure.  As such, a raft of other techniques have been developed to generate a more open pore architecture, for example, solvent casting/particulate leaching, emulsification/freeze-drying, phase separation, 3D printing (see \cite{khang} for a review of these methods) and supercritical CO$_2$ foaming \cite{barry}.  Each method has its own merits and limitations in terms of the level of pore connectivity produced, the control over the pore size, the involvement of organic solvents, and the overall economy and efficiency.  We will consider scaffolds formed by the supercritical CO$_2$ method.  This method produces scaffolds with a relatively interconnected, open-cell structure with the advantage that this can be achieved using relatively low temperatures and without organic solvents \cite{white}.  These merits make these scaffolds amenable to the incorporation of biological materials such as growth factors \cite{davies} and even mammalian cells \cite{ginty}.

The use of post-processing techniques to further engineer the structural properties of these foams would strongly support these methods, for example, to further improve the pore connectivity and fluid transport (essential for cell ingress and nutrient perfusion), or to provide a level of fine-tuning the structure towards individual cell and tissue types.  A further challenge posed by PLA-based scaffolds is the polymer's hydrophobicitiy which strongly inhibits the uptake of water-based fluids, such as cell culture media.  The use of a pre-wetting stage with ethanol has been shown to enhance the final uptake of water into hydrophobic scaffolds \cite{mikos}.  Further strategies to overcome the hydrophobicity and improve cell penetration within these polymers include the use of suitable co-polymers \cite{oh} and surface coatings \cite{intranuovo}.

High power ultrasound finds diverse applications, from cleaning and homogenizing to chemical synthesis to sterilization \cite{leighton,leong}.  High power ultrasound refers to sound waves in the ultrasonic range (frequencies greater than approximately 20 kHz) that are of high power (typically 50 W and above).  These waves generate intense local agitation of the ambient fluid.  Mostly this occurs through cavitating bubbles which collapse and generate intense pressure and temperatures on the micro-scale. 
The ability of high power ultrasound to open up the scaffold structure was first demonstrated by Wang {\it et al} \cite{wang}.  There, exposure of 3D scaffolds with initially closed pore structure, formed via solid state foaming, to high power ultrasound led to a marked increase in pore interconnectivity and generation of an open pore structure.   This work was extended in Ref. \cite{wang2} where it was found that the enhancement in pore interconnectivity and permeability increases with temperature, pore size and ultrasound power.  Guo {\it et al.} \cite{guo} applied high power ultrasound to solid-state fabricated PLA foamed sheets and noted a similar increase in pore interconnectivity.   Lee {\it et al.} \cite{lee} exposed thin electrospun scaffolds composed of PLLA to high power ultrasound and observed around a 15\% increase in porosity.  This study went on to seed cells on the scaffolds and observed the cell infilitration to increase strongly in insonified scaffolds.

Here we further examine the capacity of high power ultrasound to modify the internal structure and transport properties of {\it thick} foam polymer scaffolds.  Where Wang {\it et al} \cite{wang,wang2} and Guo {\it et al} \cite{guo} considered solid-state foams, our focus is on scaffolds formed via supercritical CO$_2$ foaming.  This type of scaffold is at a more advanced stage in biomedical research, having successfuly demonstrated the controlled release of proteins \cite{howdle2001} and \cite{ginty2008}, incorporation of mammalian cells \cite{ginty}, promotion of bone formation \cite{kanczler2010} and the induction of angiogenesis in vitro \cite{kanczler2007}.  The modification of scaffold structure is analysed via micro x-ray computed tomography, scanning electron microscopy and acoustic microscopy.  We pay particular attention to how insonification improves fluid transport through the scaffold, of essential importance 
for cell infusion and during tissue growth, by monitoring the uptake of water into the scaffold.  The effect of pre-wetting the scaffold with ethanol (a means to aid overcoming the hydrophobicity of PLA) is also studied.

\section{Materials and methods}
\subsection{ PLA scaffolds formed via supercritical CO2 foaming}

The scaffolds were composed of PLA (Purac, Gorinchem, Netherlands), with a density of $1200 ~$kg m$^{-3}$ (manufacturer's specification) and molecular weight (weight averaged) of $55$ kDa (determined in our laboratories via NMR).  The foamed scaffolds were fabricated using the supercritical CO$_2$ method as detailed in \cite{white}.  In brief, granular polymer was weighed into each well of a Teflon multi-well mould.  The mould was placed inside a 60 ml high pressure autoclave which was heated to, and maintained at, $35^{\rm o}$C.  Compressed CO$_2$ is then introduced, maintaining a pressure of 230 bar.  The vessel was later depressurized (at a constant rate) to ambient pressure.  The porous scaffolds fabricated had diameters of approximately 10 mm and were 5--10 mm in height.  A non-porous skin was removed by a scalpel blade prior to immersion and ultrasound treatment. 

\subsection{Scafffold immersion and treatment}
Each scaffold was immersed in approximately 3 cm of liquid (water or ethanol) within a test-tube and weighed down (since the scaffold is initially buoyant) by a piece of rubber tubing.   The test tube was cooled in a water-bath maintained at $5^{\rm o}$C to avoid reaching the glass transition temperature of the polymer.  The natural glass transition temperature of PLA lies in the range 30-60$^{\rm o}$C, while the presence of ethanol (which acts as a plasticizer) can reduce this  to around $10^{\rm o}$C \cite{ahmed,parker1,parker3}.  

Scaffolds were subjected to a treatment of high power ultrasound.  We employed a commercial ultrasound sonicator (Hielscher UP 100, maximum power output 100W) at, unless otherwise stated, a power level of 20 W and duty-cycle of $20\%$.  The set-up for sonicating the scaffolds is illustrated in Figure 1 .  The sonotrode was inserted into the top of the rubber tubing and a gap of $\sim$ 1 cm maintained to the scaffold (as direct contact could  lead to rapid erosion of the scaffold).  

\begin{figure}
\centering
\includegraphics[width=0.4\textwidth]{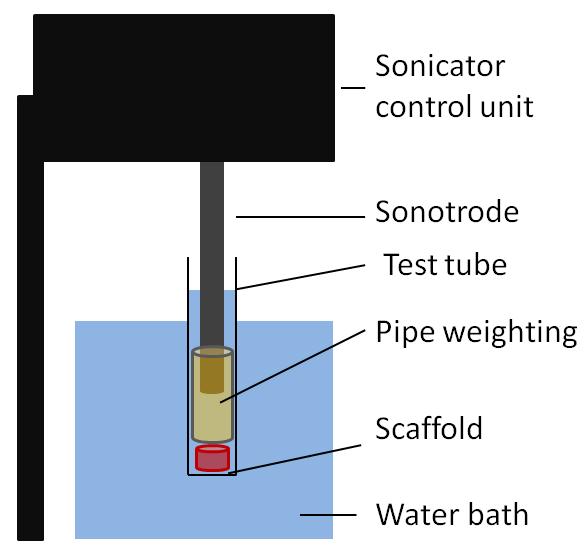}
\caption{Illustration of the set-up for sonicating the scaffolds.}
\end{figure}

We assessed three power ultrasound treatment strategies described below.    Sonicator tips shed traces of metal (titanium) during operation, and so in each case  the water was replaced every 30 minutes to minimize contamination. 
\begin{itemize}
\item {\it Mild protocol}: The scaffold is immersed in water and exposed to ultrasound of moderate power (sonicator settings of 20 W and a duty cycle of $20\%$). 
\item {\it Extreme protocol}: Again, the scaffold is immersed in water.  The sonicator was set at its greatest power output of 100 W and a duty cycle of 100\% (continuous operation). 
\item {\it Mild-with-prewetting protocol}: This protocol is the same as the mild protocol but where the scaffold was first sonicated in pure ethanol before the immersion fluid was changed to pure water and sonication continued.  
\end{itemize}

\subsection{Measurement of fluid uptake}
To monitor the uptake of fluid into each scaffold we measured its mass and volume, the latter required to allow for expansion or contraction of the scaffold over time (in practice such volume modifications were negligible).  The mass was determined using a precision balance (Mettler Toledo AB204-S) with an accuracy of 0.1 mg.  The scaffold maintained its cylindrical shape throughout the wetting process and so we estimated its volume from diameter and height measurements using calipers with a precision of $10~\mu$m.   Finally, the scaffolds were air-dried and weighed so as to assess any mass loss during the ultrasonic treatment.  

\subsection{Filling fraction}
To monitor the filling of the scaffold pores with fluid during immersion we define a filling fraction $F(t)$,
\begin{equation}
F(t)=\frac{V_{\rm fluid}(t)}{V_{\rm pore}(t)},
\end{equation}
where $V_{\rm fluid}(t)$ and $V_{\rm pore}(t)$ are the volumes of fluid and pore space, respectively, within the scaffold.  
$F=0$ corresponds to when the scaffold is free from fluid; $F=1$ corresponds to complete saturation.  We define $t=0$ to be the start of the immersion, i.e. $F(t=0)=0$.  The fluid volume within the scaffold $V_{\rm fluid}(t)$ is derived from the increase in total scaffold mass via $V_{\rm fluid}(t)=[m_{\rm tot}(t)-m_{\rm tot}(t=0)]/\rho_{\rm fluid}$.   We take $\rho_{\rm fluid}=998$ kg m$^{-3}$ for water  and $\rho_{\rm fluid}=789$ kg m$^{-3}$ for ethanol \cite{lide}. 
%While these values correspond to $20^o$C, over the temperature range in the experiments ($0-30^o$C), the densities changed by less than 0.5\% and so can be neglected.  

The pore volume is determined from $V_{\rm pore}(t)=V_{\rm tot}(t)-V_{\rm frame}$, where we assume the total volume to vary but the frame volume to be fixed (we will see that the frame loss is small enough to be neglected).  Combining these, and using the relation $V_{\rm frame}=m_{\rm tot}(t=0)/\rho_{\rm frame}$, we arrive at,
\begin{equation}
F(t)=\frac{\rho_{\rm frame}}{\rho_{\rm fluid}}\left(\frac{m_{\rm tot}(t)-m_{\rm tot}(t=0)}{V_{\rm tot}(t)\rho_{\rm frame}-m_{\rm tot}(t=0)} \right).
\end{equation}
In the two-stage (ethanol-then-water) wetting protocol, it was useful to follow the replacement of ethanol by water.  Assuming that the scaffold is completely saturated with ethanol and that its total volume is fixed,  then the fluid volume must constant and any changes in the total scaffold mass must be due to a change in the density of the fluid within.  An expression for the time-dependent fluid density $\rho_{\rm fluid}(t)$ is obtained by setting $F=1$ in Equation (2) and rearranging,
\begin{equation}
\rho_{\rm fluid}(t)=\rho_{\rm frame} \left(\frac{m_{\rm tot}(t)-m_{\rm tot}(t=0) }{V_{\rm tot}\rho_{\rm frame}-m_{\rm tot}(t=0)}\right).
\end{equation}
To monitor fluid uptake via these equations, the scaffolds were removed at regular time intervals and their mass and dimensions recorded.  Excess water bound to the scaffold due to surface tension, as well as deviations of the scaffold from a cylindrical shape, introduce systematic errors in $F$ and $\rho_{\rm fluid}$ which may reach 3\%.  

\subsection{Micro x-ray-computed tomography}
Treated and control scaffolds were characterised by micro x-ray-computed tomography (SkyScan 1174, SkyScan, Aartselaar, Belgium) so as to extract the mean pore size.  Measurements were obtained at a voltage of $50$ kV, current of $800 \mu$A and voxel resolution of $11.9 \mu$m.  The transmission images were reconstructed using the SkyScan supplied software (NRecon). The mean pore size was obtained using direct morphometric calculations in the SkyScan CTAn software package. 

\subsection{Scanning electron microscopy}

Treated and control scaffolds were dissected with a scalpel to reveal an inner surface which was imaged using scanning electron microscopy (SEM; JEOL JMS-6060 LV, JEOL Ltd., Welwyn Garden City, Hertfordshire, UK). The scaffolds were sputter coated with a thin layer of gold (Balzers Union SCD 030, Balzers Union Ltd., Liechtenstein) before being imaged (at 10kV) with the associated Smile View program (JEOL Ltd., Welwyn Garden City, Hertfordshire, UK).  This approach enabled detailed visual inspection of the scaffold pore structure before and after ultrasound treatment.

\subsection{Acoustic microscopy}
Treated and control scaffolds were imaged via an in-house acoustic microscope \cite{parker2} and these images analysed to provide a measure of porosity. This approach was employed as  an additional means to characterize the pore structure, but also as a proof-of-principle demonstration of acoustic propagation through the scaffold (made possible by the high level of saturation achieved \cite{parker3}).  The microscope directs focussed pulses of sound, through water, to the specimen and detects the  time-gated, back scattered signal, outputting a voltage trace.  By moving the pulsing/receiving transducer unit above the sample on a motorized position system, we obtained a 3D map of the back scattered signal in the form of a voltage. The microscope operates at 100 MHz and has a focal distance of 6 microns. The lateral and axial resolution are 25 and 40 microns, respectively \cite{parker2}.

Each scaffold was dissected vertically to expose an internal surface.  $C$-scan images were taken on three 2mm-square regions of the exposed surface (located at the centre and opposing sides of the exposed surface), with the focal plane of the microscope aligned with the surface plane. The acoustic $C$-scans represent a two-dimensional map of pore space (no reflection) versus frame (non-zero reflection).  

We parameterized the local porosity via image analysis.  $C$-scan images were converted to a binary image by setting a threshold intensity (taken to be 0.1 times the peak, as described in Appendix A).  This separates regions of the image representing frame (ascribed a value of 0) from those with of pore space (ascribed a value of 1). The average image intensity then represents the ratio of pore area to non-pore area, denoted ‘2D porosity’.  Assuming that the pore structure is sufficiently isotropic and homogeneous then the 2D porosity should map on to the 3D porosity.  

\section{Results}
\subsection{Fluid uptake}
\subsubsection{Fluid uptake under the mild protocol}

\begin{figure}[t]
\centering
\includegraphics[width=0.7\textwidth]{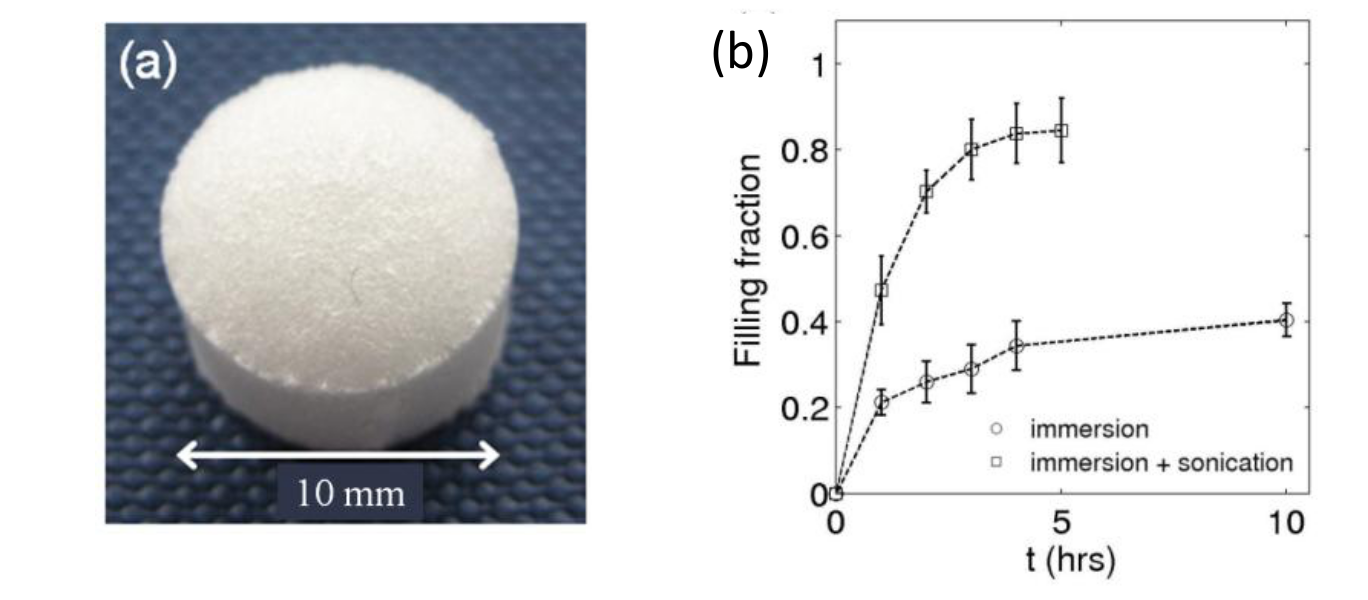}
\caption{(a) Representative image of a supercritical CO$_2$ fabricated scaffold.  (b) Water uptake in the scaffolds under immersion (circles) and immersion plus the mild protocol of ultrasound (squares).  Error bars represent standard deviation of measurements on 4 scaffolds.  }
\end{figure}
The external appearance of a representative scaffold is shown in Figure 2(a).  In Figure 2(b) we present results on the fluid uptake in the scaffolds under immersion in water (circles) and the mild protocol of ultrasound exposure (squares).  Under immersion alone, the filling fraction saturates after several hours to only $(40\pm4)\%$.  Exposure to the mild protocol of ultrasound dramatically improves the fluid uptake, which rises to approximately $(83\pm7)\%$.  Incidentally, exposure of the scaffolds to a standard laboratory sonic bath (data not presented) caused little improvement in the filling fraction, indicating that the high power sonicator is particularly effective for this purpose.  However, this was still not sufficient to promote full fluid uptake in the scaffolds.

\subsubsection{ Fluid uptake under the mild-with-prewetting protocol}
In an effort to obtain 100\% water uptake into the scaffolds, we employed a pre-wetting stage with ethanol.  In a previous study involving immersion \cite{mikos} this was shown to improve the uptake of water.  Direct immersion of the scaffolds in ethanol (circles in Figure 3(a)) led to a more {\it rapid} fluid uptake than seen previously with water but the filling fraction again saturates at around $40\%$.  This indicates that restricted pore interconnectivity, rather than the polymer hydrophobicity, is the main barrier to full fluid uptake.  A visual confirmation of this is shown in Figure 3(b)(top): under immersion in ethanol the scaffold becomes translucent, revealing a network of trapped air pockets.  
Application of high power ultrasound (squares in Figure 3(a)) had a dramatic effect in this case, raising the filling fraction to $(95\pm 8)\%$.  Indeed, the trapped air pockets became removed from the scaffold, as shown in Figure 3(b) (bottom). In addition, the fluid uptake occurred rapidly, within approximately 1 hour of insonification.  

\begin{figure}[b]
\centering
\includegraphics[width=0.9\textwidth]{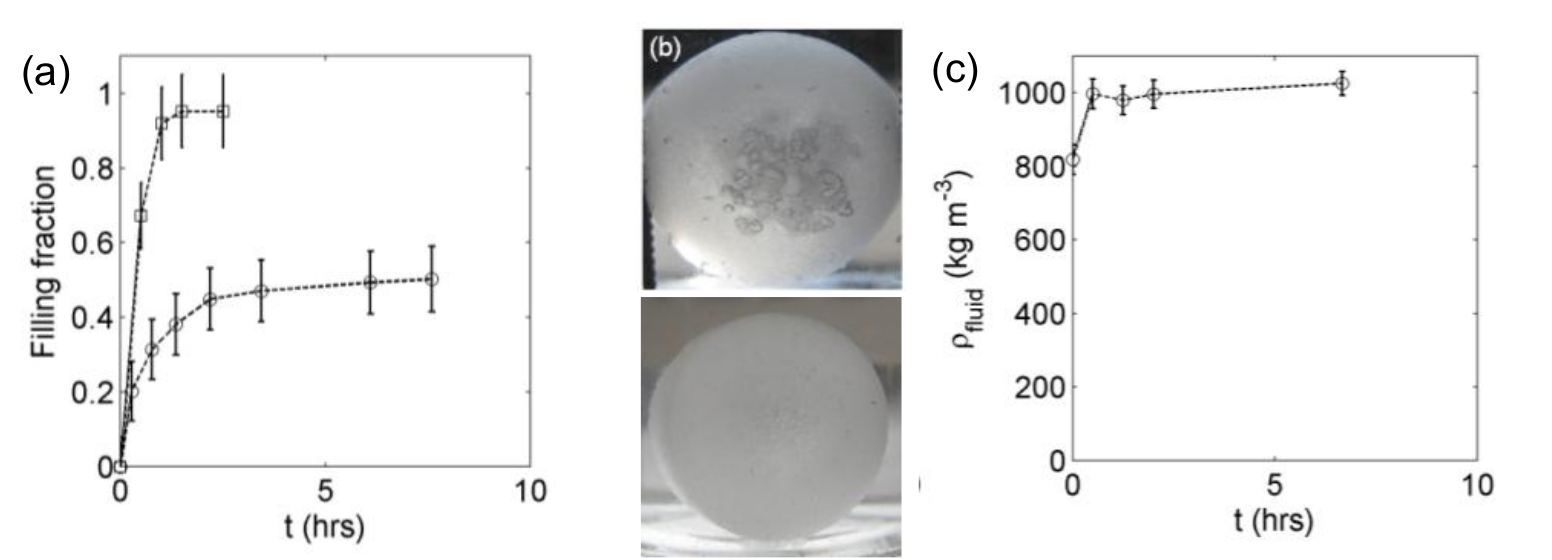}
\caption{(a) Ethanol uptake under immersion (circles) and with mild protocol of ultrasound (squares). (b) Image of the ethanol-laden scaffold with bubbles/before sonication (top) and without bubbles/after sonification (below). (c)  Fluid density in the scaffold following replacement of the immersion fluid with water. Error bars represent standard deviation over measurements on 4 scaffolds.   }
\end{figure}

Following the complete uptake of ethanol into the scaffolds, the immersion fluid was replaced with water.  Over time the density of the fluid increased from that of ethanol to that of water, as shown in Figure 3(c).  During this time, the scaffold sunk, confirming the transfer of fluids within the scaffold.  This fluid exchange appearred to be completed after approximately 1 hour.

\subsubsection{Fluid uptake under the extreme protocol}
The scaffolds were exposed to the extreme ultrasound protocol (100W, duty cycle 100\%).  As shown in Figure 4, although the uptake of water is slow, there was a significant improvement in overall uptake compared to the mild ultrasound protocol (20W, duty cycle 20\%), here reaching $(93\pm 5)\%$ filling.  

\subsubsection{Fluid uptake via the mild-with-prewetting protocol versus the extreme protocol}
We applied the mild-with-prewetting protocol over a fixed time, consisting of 2 hours of insonification in ethanol followed by 2 hours insonification in water) to a group of identically-fabricated scaffolds.  Similarly, we applied the extreme protocol for a fixed time of 10 hours to another group of scaffolds.  The final fluid uptake achieved is summarised in Table 2.  The mild treatment led to 100\% filling in all cases (within measurement error).  In contrast, the extreme treatment led to an average of 87\% filling.  
The mass loss from the scaffolds during mild treatment was on average 0.6\%.  Under the extreme treatment, the mass loss was larger, being 1.9\% on average.

\begin{figure}
\centering
\includegraphics[width=0.5\textwidth]{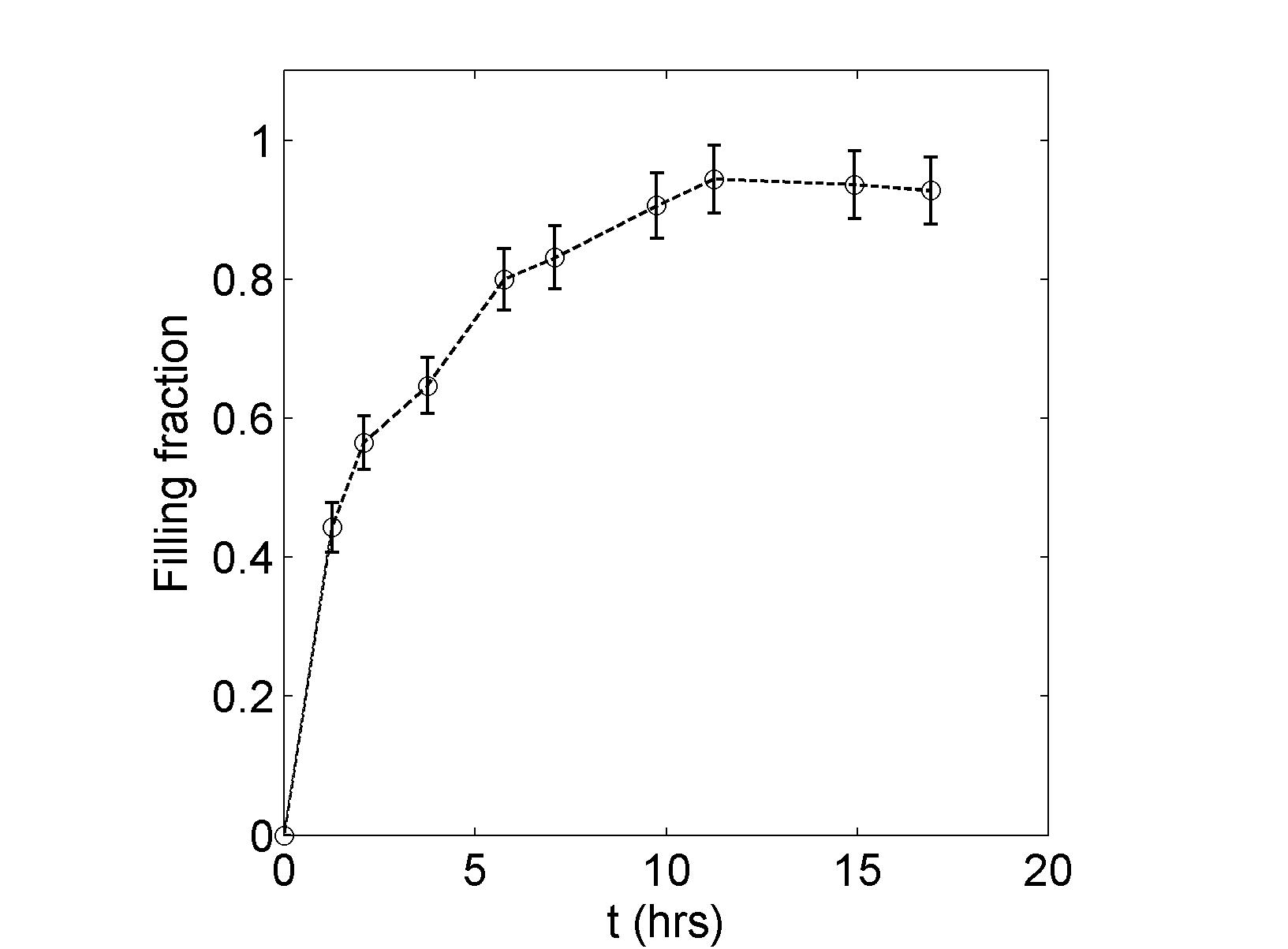}
\caption{Uptake of water under exposure to the extreme protocol of ultrasound. Data represents the mean and standard deviation of measurements of 4 scafffolds.}
\end{figure}

\begin{table}
\centering
\begin{tabular}{| l | c | c |}
  \hline                        
   & Mild-with-prewetting protocol & Extreme protocol \\
\hline
 Final filling fraction (\%) & $101 \pm 3$ & $87\pm5$ \\
Average percentage mass loss (\%) & 0.6 & 1.9 \\
  \hline  
\end{tabular}
\caption{Scaffold characteristics after treatment with the mild-with-prewetting protocol (2 hours in ethanol followed by 2 hours in water) and extreme protocol (10 hours in water).  Data represents the mean and standard deviation of measurements of 8 scafffolds.}
\end{table}

\subsection{Effect of sonication on pore diameter and structure}
\begin{figure}[b]
\centering
\includegraphics[width=0.8\textwidth]{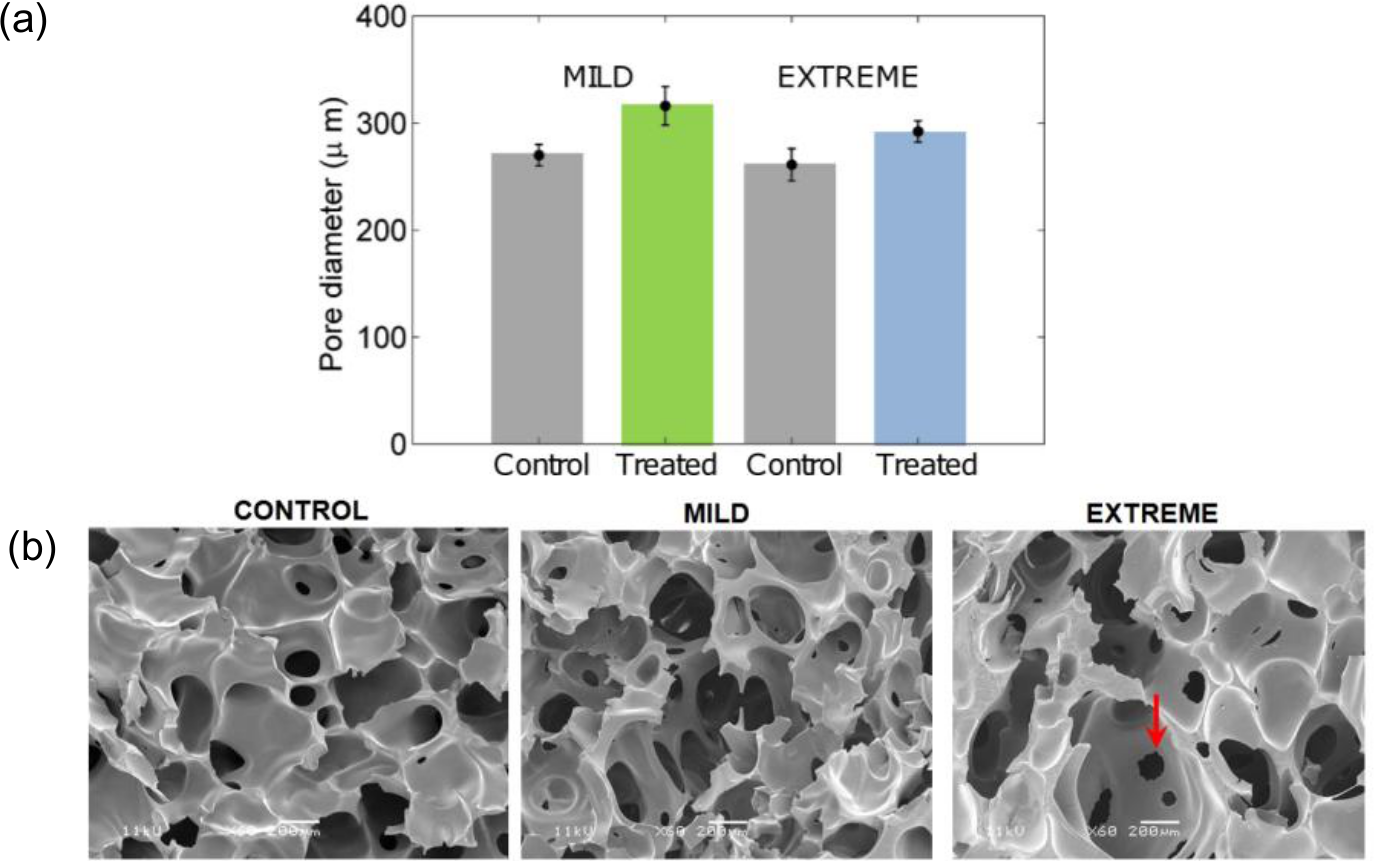}
\caption{(a) Mean pore diameter (over 4 scaffolds) following the mild-with-prewetting and extreme protocols of ultrasonic treatment (detailed in Section 2.2), compared with controls (error bars represent standard error). (b) Representative SEM images of a control scaffold and scaffolds having undergone the mild-with-prewetting and extreme treatments.  Each image corresponds to a region 2mm $\times$ 16mm.  The arrow in the right hand figure highlights one of the small holes referred to in the text.}
\end{figure}Four scaffolds that had undergone the mild-with-prewetting protocol for a fixed time (2 hours in ethanol and 2 hours in water), four scaffolds having undergone the extreme protocol for a fixed time (10 hours), and four untreated control scaffolds (all from the same scaffold fabrication batch) had their pore diameters measured using micro-CT.  The results are presented in Figure 5(a).  The mild protocol with prewetting led to an increase in the average pore diameter from 270 to 315 $\mu$m (a 17\% increase) whereas the extreme protocol led to a change in average pore diameter from 261 to 292 $\mu$m (a 12\% increase).  Figure 5(b) shows SEM images of the scaffolds.  After mild treatment the pore structure was more open than the control, with large sections of pore wall having been removed.  In contrast, extreme treatment maintains the same large-scale structure as the control but with some small holes appearing in the pore walls.

\subsection{Acoustic images and analysis }
Acoustic microscopy was used to image scaffolds which had undergone the mild-with-prewetting protocol and the extreme protocol, all for the same fixed time as above, as well as a control scaffold.  Figure 6(a) displays typical $C$-scans for these three cases. 
\begin{figure}[b]
\centering
\includegraphics[width=0.75\textwidth]{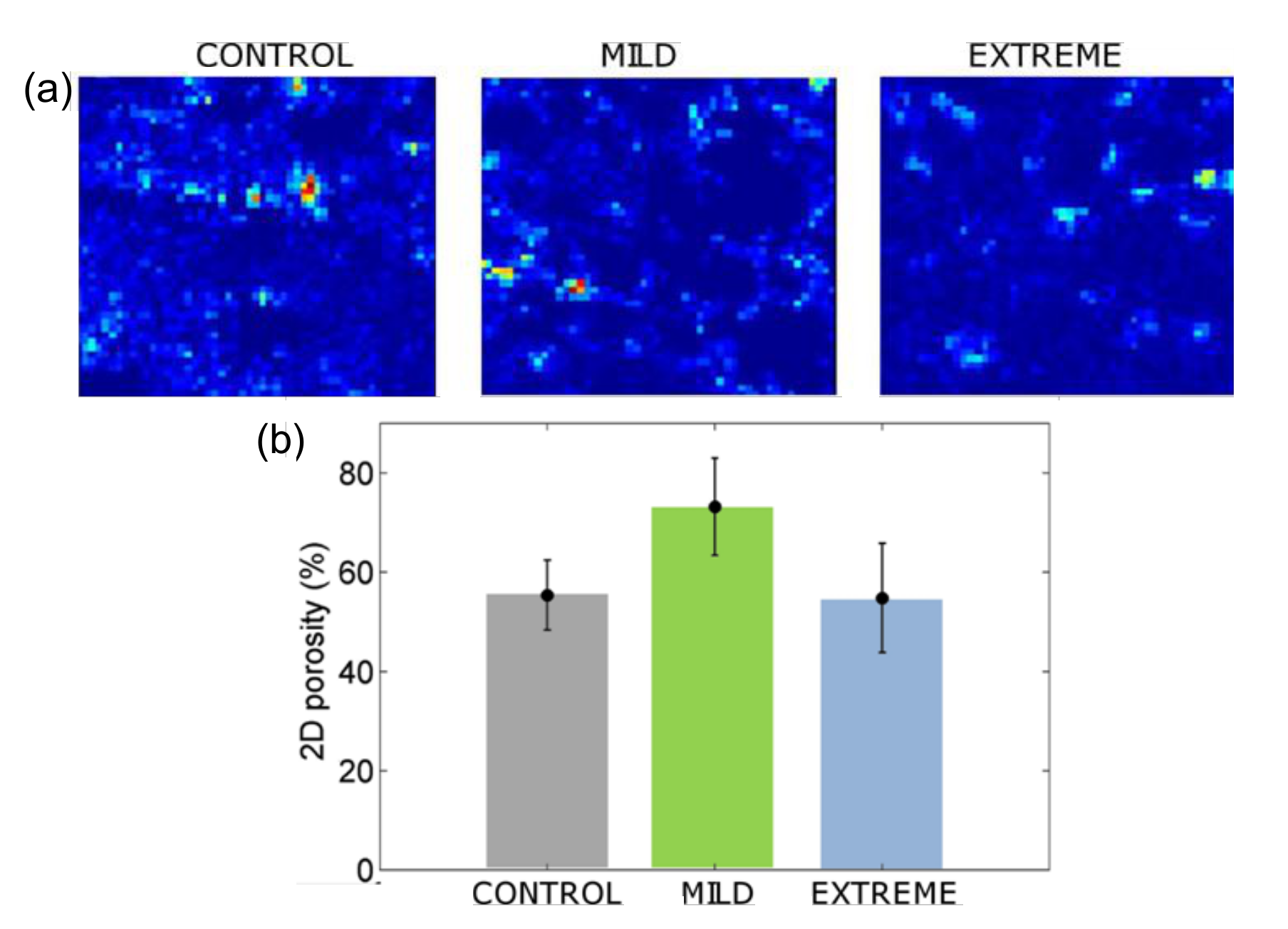}
\caption{(a) Representative $C$-scan images of the control (left), extreme-treated (right) and mild-treated (middle) scaffolds.  Each image corresponds to a 2mm-square region.  The colour amplitude corresponds to the normalized received voltage with dark blue (zero voltage) representing pore and lighter colours (including yellow to red) representing polymer scaffold. (b) Mean 2D porosity of across 3 $C$-scans performed on each scaffold (error bars represent standard error).}
\end{figure}

Three $C$-scans were recorded for each scaffold and the average 2D porosity, and its standard deviation, are presented in Figure 6(b).  The values of porosity are lower than the physical porosity in the scaffolds (anticipated to be 70-80\%).  This is due to the inherent rescaling introduced by different contrast mechanisms when used for porosity measurements \cite{mather}.  Nonetheless, for a consistent imaging type, the {\it relative} porosity values are meaningful and are of relevance here.    The results show that the 2D porosity in the control and extreme treated scaffold are very similar whereas the porosity in the scaffold treated with the mild-with-prewetting treatment regime is around 20\% higher. This is consistent with our findings in Section 3.2 that the greatest changes occurs under the mild-with-prewetting protocol rather than the extreme protocol.

\section{Discussion}
Pore interconnectivity, pore size and fluid uptake/transport are key control parameters for the successful cultivation of tissue within foam scaffolds.  Here we set out to examine the capability of high power ultrasound to modify and enhance these parameters in thick pre-processed foam tissue scaffolds.  We dedicated considerable attention to the efficacy of fluid uptake into the scaffolds induced by insonification, a practical measure of scaffold permeability and of direct importance for cell infiltration and fluid transport during tissue growth.   Such fluid uptake is usually problematic due to the restrictive, tortuous pore architecture and the hydrophobicity of the polymer.  

To date, such exploitation of high power ultrasound has been little studied.  Wang {\it et al.} \cite{wang,wang2} and Guo {it et al.} \cite{guo} found that the application of high power ultrasound to thick 3D PLA scaffolds with initially closed pore structure formed via solid-state foaming led to a marked increase in pore interconnectivity and permeability, albeit with little increase in porosity.  Lee {\it et al.} \cite{lee} performed a similar analysis for thin, electrospun nanofibre PLLA scaffolds and observed a significant increase in porosity, pore size and also cell infiltration into the scaffold.     Here we considered a distinct type of thick scaffold, with is currently in use for tissue engineering research --  super-critical CO$_2$-foamed scaffolds composed of PLA.

First, as a base line, we considered fluid uptake under direct immersion in water and found only circa 40\% pore filling.  We have additionally performed the same analysis for PLA scaffolds formed via the solvent cast/particulate leaching method (data not presented) and find circa 80\% filling under immersion.  This fabrication technique generates scaffolds with high pore interconnectivity \cite{khang}, and so the sub-optimal filling can be attributed to the hydrophobicity of PLA, inhibiting water transport through small, tortuous pathways.  Thus the even reduced filling obtained in the super-critical CO$_2$-foamed scaffolds can be attributed to the reduced pore interconnectivity in this scaffold type \cite{white}.

The application of high power ultrasound caused a significant improvement in fluid uptake of water, with around  90\% filling  achieved under the extreme protocol (100 W, duty cycle 100\%).  However, one cannot robustly achieve 100\% filling of these scaffolds with water-based sonication alone.

By pre-wetting with ethanol, followed by infusion of water which then replaces the ethanol and then exposure to high power ultrasound, we can routinely achieve 100\% filling of the scaffolds.  Our observations lead us to specify an efficient and fast method to achieve $\sim 100\%$ filling of the scaffolds with aqueous solution: 2 hours mild sonication (20W, 20\% duty cycle) in ethanol followed by 2 hours further sonication in water (conducted within a chilled environment to avoid exceeding the polymer glass transition).  

The ultrasonic treatment leads to desirable modifications of the scaffold structure.  At maximal power (the {\it extreme protocol}), the ultrasound opens up the pore structure by punching small holes in closed pore walls, in accord with observation by others \cite{wang,guo}.  This is likely due to the formation of cavitating bubbles, which are known to collapse and generate huge forces on a microscopic scale \cite{leighton}.  An increase in the mean pore size by around 12\% is observed following extreme ultrasonic treatment.  This value is comparable to that observed elsewhere using distinct scaffold types \cite{wang,wang2,guo}.

A more effective means to increase fluid transport and pore diameter is provided by mild exposure to ultrasound (20W, duty cycle 100\%) coupled with a pre-wetting stage with ethanol (the {\it mild-with-prewetting protocol}).  Here mean pore size is observed to increase by $17\%$.  The structural changes suggest that the ethanol and ultrasound combine to flush out obstructive parts of the pore framework, which may include whole pore walls.  One scenario for this effect may be as follows.  It is known that ethanol acts as a plasticizer to PLA, vastly reducing its glass transition temperature \cite{ahmed,parker1}.  Cavitating bubbles generated via the ultrasonic treatment cause local heating in the scaffold, which in turn soften the polymer and enable a more efficient structural rearrangement.  Importantly, the cavitation will be greatest in regions which obstruct and constrict the sound propagation, i.e. the regions of the scaffold architecture which we most wish to open up.  This highly targeted nature of the ultrasound is consistent with the retention of overall scaffold integrity and negligible mass loss, despite the marked increase in fluid uptake, porosity and pore diameter.  

With such highly saturated scaffolds, it becomes possible to propagate ultrasound waves throughout the scaffolds.  We demonstrated this capability through the use of acoustic microscopy to image the scaffold pore structure, with results in qualitative agreement with microCT data.  Given its capacity for non-destructive and non-invasive imaging, ultrasound may hold potential for characterising scaffolds, prior and even during tissue growth.

\section{Conclusion}

We have studied the effect of exposing thick polymer foam tissue scaffolds to high power ultrasound.  The novelty of this study lies in the use of tissue scaffolds fabricated via the supercritical CO$_2$ method, the focus on fluid uptake and transport through the scaffolds, and the inclusion of a pre-wetting stage with ethanol. The ultrasonic treatment leads to an increase in the mean pore size by approximately $10-20\%$.  More striking is the enhancement of fluid transport and pore interconnectivity in the scaffold, for which we can routinely achieve 100\% filling of the scaffolds with water (over a timescale of a few hours), overcoming the polymer hydrophobicity and partially closed pore architectures.  The ultrasound treatment works in a highly targetted manner, with no loss of scaffold integrity and negligible polymer loss.  These effects are optimized when the ultrasound treatment is coupled to a pretting stage with ethanol.  These capabilities may provide a useful and economical tool for optimizing scaffold properties post-fabrication for specific tissue engineering purposes.  Furthermore, given the demonstrated capacity to achieve ~100\% filling of PLA scaffolds, it becomes possible to propagate sound waves throughout these thick scaffolds.  We hope in future to explore the use of low-power ultrasound to characterise and image the internal structure of the scaffolds, both in isolation and during tissue growth.

\ack

We thank Dr Mel Holmes (University of Leeds) and Dr Melissa Mather (University of Nottingham) for discussions, and the Biotechnology and Biology Science Research Council for funding (BBSRC ref: BB/F004923/1).

\appendix
\section{Thresholding in the acoustic image processing}
The image processing discussed in Section 2.7 requires the setting of a threshold in order to extract the binary image from the analog image.  We analysed a single scaffold image and determined its 2D porosity for varying values of this threshold value, with the results shown in Figure A1.  For very low thresholds, where voltage noise becomes significant, we measure anomalously low 2D porosity.  For high thresholds ( $> 0.7$) we see anomalously high thresholds, caused by the effective removal of all but the strongest reflections.  In between we see a large, well-behaved range ($0.05 <$ threshold $< 0.7$) where the 2D porosity is insensitive to the threshold value and it can be assumed that the method successfully captures the porosity value.  Based on this, we employed a threshold of 0.1 in this work.

\begin{figure}[b]
\centering
\includegraphics[width=0.5\textwidth]{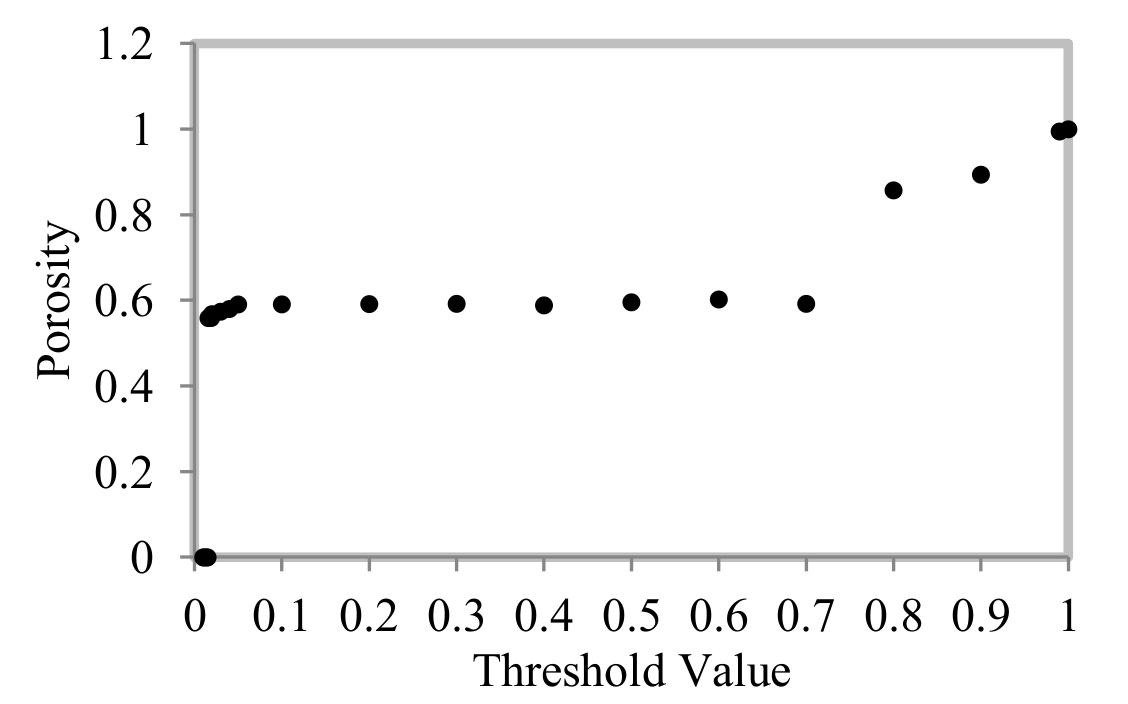}
\caption{2D porosity of a typical C-scan image against the intensity threshold value used in the binary image generation.}
\end{figure}

\end{document}